\documentstyle[pjw_cite3,psfig,graphicx]{mn}
\title[The X-ray eclipse of OY~Car resolved with XMM-Newton]
{The X-ray eclipse of OY~Car resolved with XMM-Newton: 
X-ray emission from the polar regions of the white dwarf}
\author[P.\,J.\,Wheatley and R.\,G.\,West]
{Peter J.\  Wheatley and Richard G.\ West \\
Department of Physics and Astronomy, University 
of Leicester, University Road, Leicester, LE1 7RH \\
}

\def\TD-1{\it TD-1\rm }
\def\tkev{\thinspace{ke\kern-.15em V}}
\def\tev{\thinspace{e\kern-.15em V}}
\def\etal{et~al. }

\begin{document}
\maketitle

\begin{abstract}
We present the XMM-Newton X-ray eclipse lightcurve of the dwarf nova OY~Car. 
The eclipse ingress and egress are well resolved for the first time in any 
dwarf nova placing
strong constraints on the size and location of the X-ray emitting region. 
We find good fits to a simple linear eclipse model, giving ingress/egress 
durations of $30\pm3$\,s ($\rm \Delta\phi_{orb}=0.0055\pm0.0006$).
Remarkably this is shorter than the ingress/egress duration of the
sharp eclipse in the optical as measured by \scite{Wood89} and 
ascribed to the white dwarf ($43\pm2$\,s). 
We also find that the X-ray eclipse is narrower than the optical
eclipse
by 14$\pm$2\,s, which is precisely the difference required to align
the second and third contact points of the X-ray and optical
eclipses. 
We discuss these results and conclude that X-ray emission in OY~Car
most likely arises from the polar regions of the white dwarf. 

Our data were originally reported by \scite{Ramsay01a}, but they did not make 
a quantitative measurement of eclipse parameters. We have also corrected 
important timing anomalies present in the data available at that time. 

\end{abstract}

\begin{keywords} Accretion, accretion disks -- Binaries: eclipsing  -- 
Stars: novae, cataclysmic variables -- Stars: individual: OY Car -- 
X-rays: stars.
\end{keywords}

\section{Introduction}
X-ray eclipses have now been observed in three dwarf novae in quiescence
\cite{Wood95,Mukai97,Teeseling97,Pratt99a}. 
The eclipses are deep, narrow, and centred on the white dwarf, 
providing strong evidence 
that the boundary layer is the source of the X-ray emission. 
The boundary layer is the region in which accretion disc
material settles from its Keplerian velocity onto the surface of the white 
dwarf, giving up some of its kinetic energy \egcite{Pringle79}. 

The most 
powerful
X-ray
eclipse study to date is of HT~Cas with ASCA \cite{Mukai97}. 
Mukai \etal made a detailed investigation 
of the eclipse depth, width and shape, and concluded that the eclipse is 
centered on the white dwarf, 
consistent with being total, 
and sufficiently sharp to limit the extent of X-ray emission to 1.15
the radius of the white dwarf. 

Eclipsing dwarf novae are quite rare, and tend to be fainter in X-rays
than their 
low-inclination counterparts \cite{Teeseling-tenCVs}, so eclipse studies to 
date have been count-rate limited. 
HT~Cas is the brightest eclipsing dwarf nova, and yet the ASCA data 
did not resolve the ingress and egress of the eclipse. 
Thus Mukai \etal were unable to measure the size or geometry of the X-ray 
emitting region.

In this paper we present the first XMM-Newton observations of an eclipsing 
dwarf nova: OY~Car. The high effective areas of the XMM EPIC 
cameras allow us to resolve the ingress/egress of an X-ray eclipse for the 
first time. 
Our data have previously been reported by \scite{Ramsay01a} and 
\scite{Ramsay01b},
but those authors did not attempt to make a quantitative measurement of
the eclipse parameters, 
partly because of important timing anomalies that were present in the
data available at that time. 
We have corrected these anomalies, and the resulting lightcurves allow
us to place tight constraints on the size and location of the X-ray 
emitting region in OY~Car. 

Quiescent X-ray eclipses were discovered in OY~Car with ROSAT by 
\scite{Pratt99a}. The eclipse is narrow, deep and centred on the white dwarf, 
but the ROSAT observation did not yield enough counts for further analysis. 

\begin{figure*}
\begin{center}
\includegraphics[height=17.7cm,angle=270]{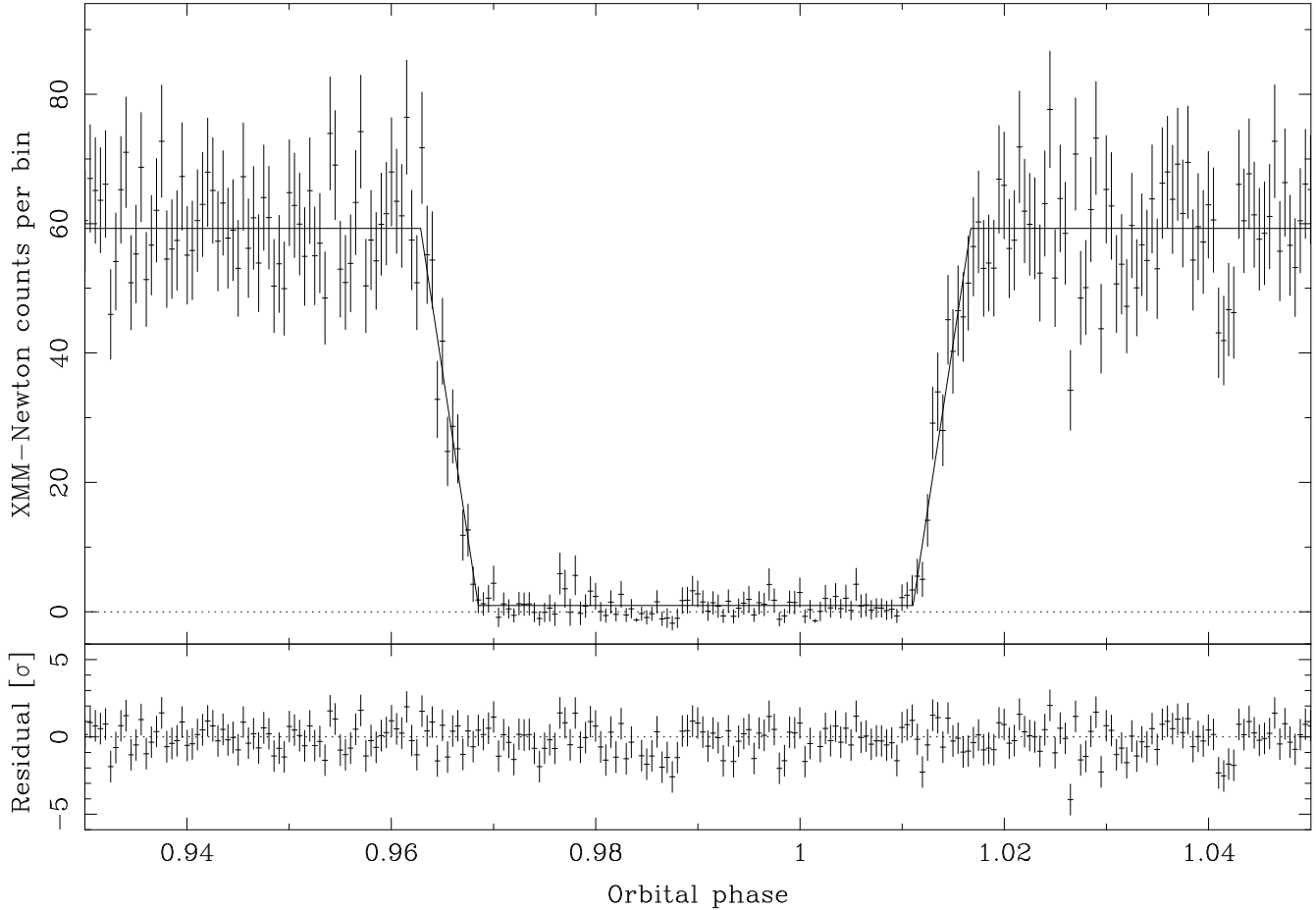}
\caption{\label{fig-fit} 
Folded XMM-Newton eclipse lightcurve of OY~Car from the June 2000
observation. Data from the 
three EPIC X-ray cameras have been co-added, binned into 2000 phase bins,
and the eclipse fitted with a simple piecewise linear model. The 
ingress/egress duration is found to be $29.5\pm_{2.4}^{2.8}$\,s, which is 
substantially shorter than the eclipse of the white dwarf observed in the 
optical/ultraviolet.}
\end{center}
\end{figure*}

\begin{figure}
\begin{center}
\includegraphics{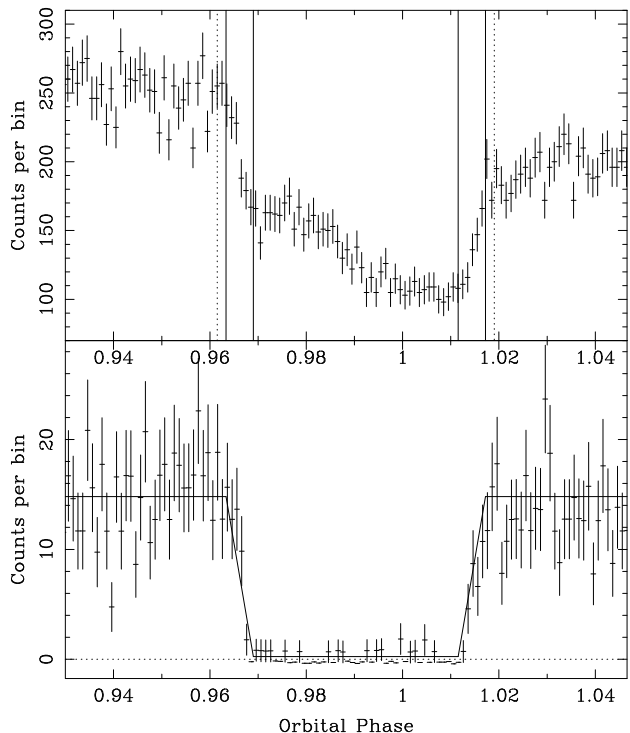}
\caption{\label{fig-obs2_om} 
Folded XMM-Newton eclipse observations of OY~Car from the second
observation in August 2000. The top panel shows a fold of 3 OM
eclipse observations with the B filter. The bottom panel shows a fold
of 3 eclipse observations with the EPIC-mos X-ray cameras and 2
eclipse observations with the EPIC-pn camera. 
The solid curve plotted over the X-ray data represents our best fit to
the June 2000 data from Fig.\,\ref{fig-fit} (scaled by count rate).
The solid vertical lines on the OM panel represent our measured X-ray
contact points, and the dotted lines represent the separation 
of the first and fourth contact points of the white dwarf as derived
by Wood et al.\  (1989) from ground-based optical observations. 
Both data sets are binned into 1000 orbital phase bins, corresponding
to 5.5\,s per bin.}
\end{center}
\end{figure}

\begin{figure}
\begin{center}
\includegraphics{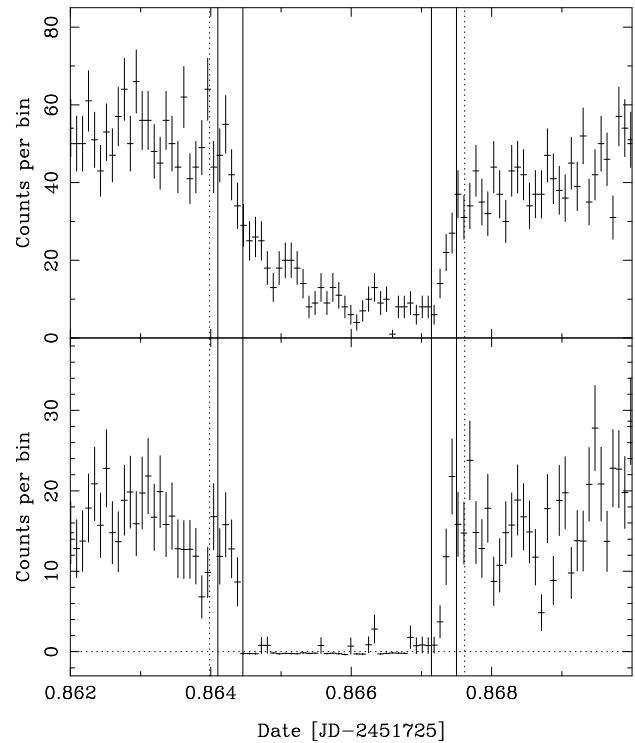}
\caption{\label{fig-obs1_om} 
Simultaneous eclipse observation of OY~Car with the UVW1 filter of the
XMM-Newton OM (top) and the EPIC X-ray cameras (bottom). 
This is the only eclipse from the first XMM-Newton observation (June
2000) for which an OM lightcurve can be recovered with full timing 
information. The solid vertical lines indicate the X-ray contact
points derived from our fit to the full folded lightcurve. 
The dotted vertical lines indicate the separation of the first and fourth
contact points of the white dwarf  derived by Wood et al.\  (1989) 
from ground-based optical observations. The bin size is 6.7\,sec in
both panels. }
\end{center}
\end{figure}

\section{Observations} 
\subsection{XMM-Newton EPIC X-ray data}
OY~Car was observed with XMM-Newton \cite{Jansen01} for 51\,ks on 2000 June 29/30 
(JD\,2\,451\,725) shortly after an outburst. 
The  data have been presented previously by \scite{Ramsay01a} but suffered 
from
two serious anomalies at that time. First, the absolute timing of all
lightcurves were unknown. Thus the relative timing between the
XMM-Newton cameras was unknown and
Ramsay \etal were forced to align the lightcurves by eye. Second, there 
was a timing anomaly in the EPIC-pn lightcurve, caused by an onboard 
counter overflow, after which all events 
were offset by $\sim$300\,s. Ramsay \etal rejected events after 
this time. 

We correct both these anomalies and combine data from 
all three EPIC cameras. Rather than extracting binned lightcurves and then
folding these to produce a mean eclipse profile, we extract event
lists for circular regions around OY~Car and produce a folded lightcurve 
by calculating orbital phase for each event. 
This avoids the need to calculate errors on
lightcurves with small numbers of counts per bin.  Source counts were
extracted from circles with 44\,arcsec radius in the MOS cameras and
60\,arcsec in the pn camera.
Background counts were collected from 44 to 88\,arcsec annulii in the
MOS cameras, and the whole of the rest of the CCD was used in the 
pn camera (except the first ten rows, which are dominated by electronic
noise). 
We produced two background-subtracted folded lightcurves, 
one with 1000 and one with 2000 phase bins. Throughout this paper we
use the ephemeris of \scite{Pratt99a} in which the orbital period of OY~Car is 
$5453.64732\pm0.00002$\,s.
The eclipse profile from the 2000-bin lightcurve is 
presented in Fig.\,\ref{fig-fit}. In total, 10 eclipses were covered
with all three EPIC cameras. It can be seen that the eclipse is
occurring around one minute earlier than predicted by the ephemeris of 
\scite{Pratt99a}. We find the centre of eclipse occurs at an orbital
phase of $0.9898\pm0.0004$. 
This difference is much greater than the stated error on
the ephemeris and we can find no mistake in our phase
calculation. XMM-Newton times are in the Terrestrial Time system (TT)
and we have applied a heliocentric correction of $+91$\,s to our
lightcurve. 

A second XMM-Newton observation of OY~Car was made on 2000 August 7 
(JD\,2\,451\,764) lasting 14\,ks. We reduced the X-ray data by the
method described above and plot the folded eclipse lightcurve in the
lower panel of Fig.\,\ref{fig-obs2_om}. Three eclipses were covered
with the two EPIC-mos cameras and two eclipses were covered with
the EPIC-pn camera. A heliocentric correction of $-23$\,s was applied to
the XMM-Newton TT event times, yielding a phase offset that is
consistent with that of the June observation. 

\subsection{XMM-Newton Optical Monitor data}
The XMM-Newton Optical Monitor (OM; \ncite{Mason01}) 
provides simultaneous optical or ultraviolet coverage of the EPIC 
field of view. The first XMM-Newton observation of OY~Car was also the 
commissioning observation for the OM {\em fast mode}, in which high time
resolution lightcurves can be accumulated. Unfortunately only two of
the ten eclipses are covered by OM fast mode exposures, and the first
of these suffers from a timing anomaly that we cannot recover. An
eclipse lightcurve has been extracted from the second exposure
and is
plotted in the upper panel of Fig.\,\ref{fig-obs1_om}. 
The observation
was made using the UVW1 filter with a bandpass of approximately 250--350\,nm. 
The lower panel of Fig.\,\ref{fig-obs1_om} shows the combined EPIC
X-ray lightcurve for the same time interval. 

By the time of the second XMM-Newton observation the OM fast mode was fully
operational and an OM lightcurve is available for all three eclipses.
The folded lightcurve 
is plotted in the
upper panel of Fig.\,\ref{fig-obs2_om}. This observation was made
using the B filter with a bandpass of approximately 380--500\,nm.

\section{Results} 
\subsection{X-ray lightcurves}
We fit the XMM-Newton eclipse lightcurve of OY~Car using a simple
piecewise linear model. Our model has five parameters: the ingress/egress 
duration, the phase of mid eclipse, the 
eclipse width (measured between mid ingress and mid 
egress), and the eclipse and out-of-eclipse count rates. 

We first applied this model to the 1000-bin lightcurve of the June
2000 observation, in which there are 
sufficient counts in all bins to allow us to apply the $\chi^2$ statistic. 
We fitted the 140 bins around mid-eclipse, and found an excellent
fit with $\chi^2$ of 135 and 135 degrees of freedom. 

The estimated errors on 
the ingress/egress duration in this fit are dominated by the large bin size 
($\approx$5.5\,s), 
and so we use the 2000-bin lightcurve for parameter estimation. 
We first fitted the 280 bins around mid-eclipse using the $\chi^2$ statistic,
in order to find an approximate fit. We did not use this fit to estimate 
errors because many eclipse bins contain too few counts. Instead we refitted 
the lightcurve using the Cash statistic \cite{Cash79}. This does not allow us 
test goodness of fit, but does allow us
to estimate allowed parameters ranges in the same way as with $\chi^2$. 

Our best Cash-statistic fit to the 2000-bin eclipse lightcurve is
plotted in Fig.\,\ref{fig-fit}. The residuals show no systematic effects 
during ingress or egress, 
and we are satisfied a linear fit is sufficient to characterise our 
eclipse data. Fitting with more complex models appears not to be warranted.  
Our best-fitting ingress/egress duration is 
$29.5\pm_{2.4}^{2.8}$\,s and the eclipse duration is $262.3\pm_{1.0}^{1.4}$\,s 
(68 per cent confidence intervals). 
The allowed ranges at 99 per cent confidence are 
25.2--36.5\,s and 259.1--266.3\,s respectively. 

\begin{figure}i
\begin{center}
\includegraphics{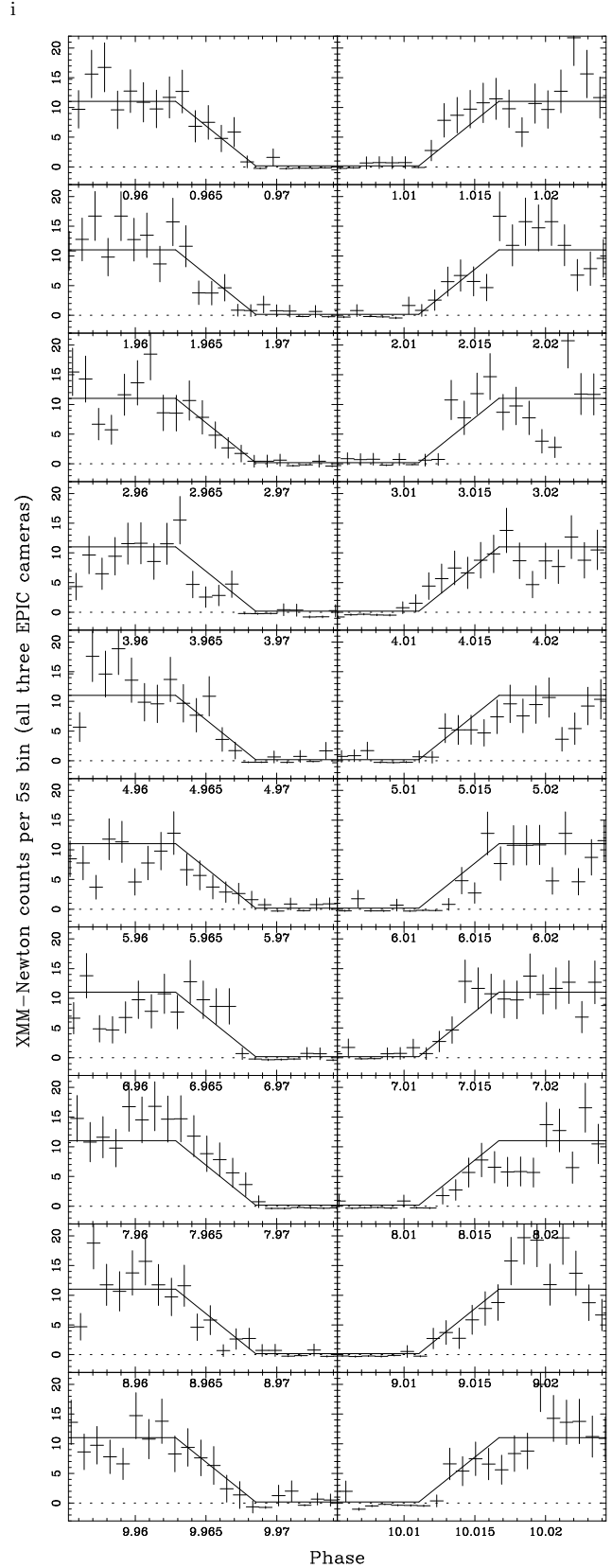}
\caption{\label{fig-all} 
The full set of eclipse ingress and egress phases from the June 2000
XMM-Newton observation of OY~Car. The solid line shows the best fit to
the full folded lightcurve plotted in Fig.\,\ref{fig-fit}. 
It is clear that the same fit is also a good representation of the
individual eclipses. }
\end{center}
\end{figure}

Figure\,\ref{fig-all} shows our best-fit model overlaid on all of the 
individual eclipse ingress and egress phases from the June 2000
observation. It is clear that our model  
is also a good representation of the individual eclipses. 

The folded X-ray lightcurve from the August 2000 observation is
plotted in the lower panel of Fig.\,\ref{fig-obs2_om}, with our best
fit to the June data overlaid (scaled by count rate). The 
August observation is shorter and the source fainter than in June, 
but it can be seen that the model fit to the June data is an excellent
representation of the August data. The X-ray eclipse lightcurve of
OY~Car seems therefore to be stable on timescales of months as well as hours. 

\subsection{Optical/UV lightcurves}
Figure\,\ref{fig-obs2_om} shows the folded B-band eclipse lightcurve
covering three eclipses during the August 2000
observation. Figure\,\ref{fig-obs1_om} shows the UVW1 eclipse
lightcurve of a single eclipse during the June 2000 observation. These
simultaneous observations confirm that the optical/UV eclipse is 
aligned with the X-ray eclipse and that the X-ray, optical and UV
eclipse ingress and egress occur on very similar timescales. However, the
low number of eclipses covered and the relatively small aperture of
the OM do not allow us to constrain the eclipse parameters as tightly
as earlier ground based observations. 

\section{Discussion} 
\subsection{Size and location of the X-ray emitting region}
For the first time we have resolved the ingress and egress of an X-ray 
eclipse in a dwarf nova. The duration and timing of these transitions 
place tight constraints on the size and location of the 
X-ray emitting region, while their shape 
constrains the distribution of X-ray emitting plasma. 

Our simple piecewise linear fit to the eclipse results in an excellent fit, 
with no systematic residuals (Fig.\,\ref{fig-fit}). We summarise our
fitted eclipse parameters in Table\,\ref{tab-ecl}, and include values
taken from the optical study of \scite{Wood89} for comparison.

Our model does not define a unique geometry, but corresponds to 
gas distributions in which an equal emission measure is occulted per 
unit time (or phase) by the limb of the secondary star. 
This is a fairly good approximation, for example,  to 
an optically-thin boundary layer
(see e.g.\  curve 3(c) in Fig.\,2 of \ncite{Wood90}). 

Remarkably our measured ingress/egress duration, $30\pm3$\,s, 
is substantially less than that 
measured in the optical by \scite{Wood89}, 
$43\pm2$\,s.
If one assumes the optical contact points represent the contact points of the 
white dwarf then the X-ray emitting region must have
a smaller extent than the white dwarf. 

Even more remarkably, our measured X-ray eclipse width, 
$262\pm1$\,s, 
is also shorter than the optical eclipse measured by
\scite{Wood89}, $276\pm2$\,s. While the ingress/egress duration
depends on the size of the emitting region, the eclipse duration is 
sensitive only to the size of occulting object. 
A narrower eclipse in X-rays implies that the secondary star must be
narrower along our line of sight to X-ray emission than it is along
our line of sight to the optical emission region. 
Since the
scale height of the atmosphere of the secondary star is negligible
with respect to the size of the white dwarf \egcite{Wood90}, this
situation can arise only if the X-ray emitting region is physically 
displaced vertically with respect to the centre-of-light in the optical. 

The difference between the eclipse durations in the X-ray and optical
bands, 14$\pm$2\,s, is precisely that required to align the second and
third contact points of the X-ray and optical lightcurves. 
Assuming the X-ray and optical eclipses are centred on each other, 
the optical eclipse must begin before and finish after the X-ray
eclipse. However, the point at which the eclipse becomes total, and
the beginning of the egress from eclipse must be aligned in the two
wavebands. This implies that, although the X-ray emitting region is vertically
displaced with respect to the optical emission of the white dwarf, the
X-ray emission does not extend beyond the limb of the white dwarf. 

\begin{figure}
\begin{center}
\includegraphics[width=8.4cm]{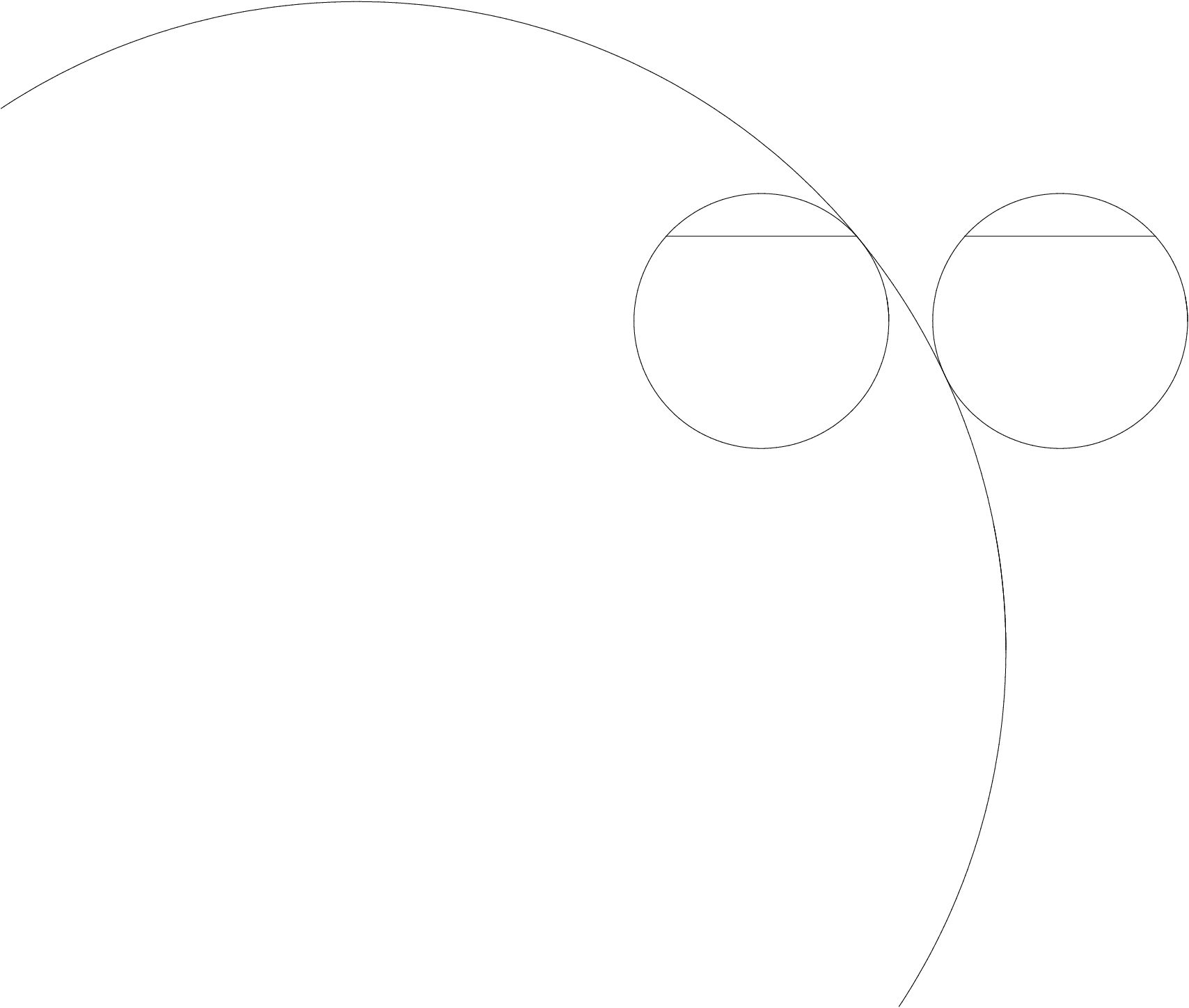}
\caption{\label{fig-diag2} 
Schematic diagram of the eclipse of the white dwarf in OY~Car, 
showing the first and second contact points of the white dwarf. 
The radius of the white dwarf has been increased by a factor three 
for clarity. 
It can be seen that the eclipse of X-rays emitted from the polar region 
will begin after the start of the optical ingress, but the end of
ingress will occur simultaneously in both wavebands. 
The same behaviour will occur in reverse during egress. 
The duration of the X-ray eclipse (measured at the half-flux points) 
will also be shorter than in the optical band. 
}
\end{center}
\end{figure}

The combination of these constraints leads us to believe that we are
seeing X-ray emission only from the upper polar region of white
dwarf. 
Figure\,\ref{fig-diag2} shows a schematic of an eclipse in which
the X-ray emission is displaced to the pole of the white dwarf.  System
parameters have been taken from \scite{Wood89}, but the radius of the 
white dwarf has been increased by a factor three for clarity. It can
be seen that such an arrangement would reproduce the implied phasings
of the X-ray and optical eclipses of OY~Car. 

\begin{table}
\begin{center}
\caption{Comparison of the eclipse parameters of OY~Car from this work
(X-rays) and from Wood et al.\ (1989; optical). $\Delta_{wi}$ and
$\Delta_{wi}$ are the durations of the eclipse ingress and egress
repectively. $\Delta\phi$ is the width of the eclipse from mid-ingress
to mid-egress. $\phi_0$ is the phase of mid eclipse. All values are
expressed in orbital phase. }
\label{tab-ecl}
\begin{tabular}{llcccc}
  & & $\Delta_{wi}$ & $\Delta_{we}$  & $\Delta\phi$ & $\phi_0$ \\\hline
This work & best fit         &\multicolumn{2}{c}{0.0054}& 0.0481 & 0.9898  \\
          & 1-$\sigma$ error &\multicolumn{2}{c}{0.0005}& 0.0003 & 0.0004  \\
          &                  &                          &        &\\
Wood et al. & mean     & 0.0077     & 0.0079      & 0.0506 & 0.0000 \\
(1989)      & rms      & 0.0004     & 0.0003      & 0.0004 & 0.0003 \\
\end{tabular}
\end{center}
\end{table}

\subsection{Origin of the vertical displacement}
X-ray emission is not usually associated with the polar regions of
the white dwarfs in dwarf novae, 
with most authors expecting emission from a narrow
equatorial boundary layer between accretion disc and white
dwarf. Emission from polar regions {\em is} seen in intermediate polars,
where a strong magnetic field truncates the accretion disc and channels the 
accretion flow onto the magnetic poles of the white dwarf. 
A similar geometry may exist in OY~Car, and this would represent the
first discovery of magnetic accretion in a classical dwarf nova. The
discovery of a 37\,min period in the XMM-Newton lightcurve by
\scite{Ramsay01a} may support this interpretation, although this is a
relatively long spin period for an intermediate polar. Alternatively we
can imagine geometries in which the entire white dwarf surface is
covered by X-ray emitting material, but that the equatorial regions
are obscured by material that is optically thick in the X-ray band
but optically thin in the optical. In both interpretations the
observed asymmetry leading to the apparent vertical displacement of
the X-ray emitting region requires that any X-ray
emission from the lower hemisphere of the white dwarf must be
obscured. This obscuration can probably be attributed to the accretion
disc itself.

\subsection{The white dwarf eclipse}
\label{sect-wd}
Our
conclusion that the X-ray emitting region has a
smaller extent than the white dwarf 
depends
critically on the assumption that the optical contact 
points measured by \scite{Wood89} accurately represent the contact points of 
the white dwarf. 
In a second paper \scite{Wood90} use realistic white dwarf and
boundary layer models to fit the eclipse shape and find the beginning
and end of ingress/egress are too gradual to be fitted with a white
dwarf filling the contact points of \scite{Wood89}, 
even with maximum limb darkening. 
As a result, their fitted white-dwarf radii (0.0144$a$--0.0163$a$) are smaller 
than that from the direct contact-point measurement (0.0182$a$) 
implying a white-dwarf ingress/egress with the same duration as we
find in X-rays. 

\scite{Wood90} discuss this discrepancy and conclude that it can be
resolved by one of two possibilities: 1) the direct method of contact
point measurement employed by \scite{Wood89} may suffer from a bias
that leads to an over-estimate of the size of the white dwarf; or
2) the range of models considered by \scite{Wood90} may not include a
model that adequately describes the intensity distribution of the
central object in OY~Car. They go on to suggest that a white dwarf model 
with a bright equatorial region that has a large extent in latitude 
might allow an acceptable fit with a 
white dwarf radius as large as that measured by \scite{Wood89}.

Given this uncertainty in the appropriate model for the central object
in OY~Car, we believe that the direct contact-point method of
\scite{Wood89} is probably a more reliable method for defining the 
contact points of the white dwarf than the fitting of \scite{Wood90}.
If one accepts 
this view then
the X-ray emission in OY~Car most probably arises 
from the polar regions of the white dwarf. 
If one does not accept this view then the extent of the X-ray emitting 
region can be as large as the white dwarf. 
However, the width of the optical eclipse found by \scite{Wood89} and
\scite{Wood90} {\em are} consistent, and so the X-ray emitting region
must still be displaced vertically with respect to the optical
centre-of-light of the white dwarf. 

\section{Conclusions}
For the first time we have resolved the X-ray ingress and egress of an 
eclipse in a dwarf nova. Comparison with high-precision
ground-based optical observations has shown that the ingress/egress duration
is substantially shorter in the X-ray band, implying the X-ray
emitting region is smaller than the white dwarf 
(although see Sect.\,\ref{sect-wd}). The eclipse
width is also smaller in the X-ray band, implying that the X-ray
emitting region must be vertically displaced with respect to the
centre of the white dwarf. The difference in ingress/egress
duration, $13\pm4$\,s, is consistent with the difference in the 
eclipse width, $14\pm2$\,s. Assuming the X-ray and optical eclipses
are centred at the same phase, this means that the second and third
contact points must be aligned. Thus, although the X-ray emission is
vertically displaced from the centre of the white dwarf, it cannot
extend beyond the limb of the white dwarf. We therefore conclude that
the X-ray emission in OY~Car most likely arises from the upper polar 
region of the white dwarf. Any emission from the lower pole must be
occulted, presumably by the accretion disc.

\section*{Acknowledgments}
We thank Janet Wood for useful discussions, and the referee Koji Mukai
for spotting a serious error in the first version of this paper.  
Research in astrophysics at the University of
Leicester is supported by PPARC rolling grants. This paper was based
on observations obtained with XMM-Newton, an ESA science mission with 
instruments and contributions directly funded by ESA Member States and the
US (NASA).

\bibliographystyle{mnras3}
\bibliography{mn_abbrev2,refs}

\end{document}